\documentclass{cflowproc}

\def\mathfont#1{\ifmmode{#1}\else{$#1$}\fi} 
\def\lae{\mathrel{<\kern-1.0em\lower0.9ex\hbox{$\sim$}}}  
\def\gae{\mathrel{>\kern-1.0em\lower0.9ex\hbox{$\sim$}}}  

\usepackage{natbib}

\begin{document}


\shortauthors{McNamara}     
\shorttitle{Star Formation} 

\title{Star Formation in Cluster Cooling Flows}   

\author{B.R. McNamara,\affilmark{1}}   

\affil{1}{Ohio University, Physics \& Astronomy, Athens, OH}   


\begin{abstract}
New X-ray observations from the {\it Chandra} and XMM-{\it Newton}
observatories have shown that cooling of the intracluster
medium is occurring at rates that are now 
approaching the star formation rates measured in cD galaxies at the bases 
of cooling flows.  Star formation proceeds
in repeated episodes, possibly indicating an intermittent fuel supply.
Coupled with new evidence for heating
by AGN, a new paradigm of self-regulated cooling and star formation in
cluster cores is emerging. 

\end{abstract}


\section{The Cooling Flow Problem}
\label{McNamara:intro}

Little has changed in our understanding of star formation in cooling
flows since the the topic was last reviewed at the 
meeting on cooling flows at Haifa in 1996.
However, the nature of the so-called ``cooling flow problem''
has changed dramatically since then.   
This controversial problem concerns the deficit
between the large cooling rates of the keV gas in
the centers of clusters and
the much smaller star formation rates observed
in central cluster galaxies located at the bases of
cooling flows (Fabian 1994).  While the cooling rates based
on {\it Einstein} and {\it ROSAT} X-ray observations (but not ASCA,
see Makishima this conference)
were claimed to be tens to hundreds of solar masses per year, 
the star formation rates
are generally between a few to several tens of solar masses
per year.  In spite of recent detections
of molecular gas in cooling flows (Edge 2001), this apparent violation of mass
continuity cannot be reconciled
by a repository of molecular gas clouds or star formation with the
Local initial mass function in central cluster galaxies.  

Adherents of the cooling flow paradigm dodged the problem by appealing to
a repository of elusive matter, such as low mass stars or a
mist of cold clouds.  Others proposed heating mechanisms (e.g.,
active nuclei, or AGN, heat conduction, 
cosmic rays, mergers, magnetic reconnection, etc.) with the potential to
inject enough energy into the
keV gas to balance radiative losses.   Nevertheless, these proposals
generally suffered various problems, such as
the need for fine tuning, but mostly
they lacked observational support.

\section{The Rumblings of a Paradigm Change}

Two recent developments have dramatically changed our view of cooling 
flows.  First, XMM-{\it Newton} grating spectra of the critical soft
X-ray band failed to detect the emission lines that dominate
cooling below 2 keV
at the predicted levels (Peterson et al. 2003, \& this conference).  
Although the spectra do not exclude cooling entirely, 
they limit the amount of gas cooling below X-ray temperatures
(where it is available to fuel star formation) to be $5-10$ 
times less than the predicted levels. 
The spectra imply that most of the cooling gas is maintained
above $\sim 2$ keV (i.e., it is being reheated),
or that it is cooling without an obvious spectroscopic signature  
(Fabian et al 2000b, Peterson et al. 2003).  Similar conclusions have been 
reached using moderate resolution CCD spectroscopy
from {\it Chandra} and XMM-{\it Newton} (McNamara et al. 2000,
David et al. 2001, 
Molendi \& Pizzolato 2001, B\"ohringer et al. 2001, Blanton et al. 2003).

Secondly, strong interactions (Carilli et al. 1994,
B\"ohringer et al. 1993) between radio sources and
the intracluster medium are now commonly seen in
{\it Chandra} images (McNamara et al. 2000, Fabian et al. 2000,
and see McNamara 2002, and Nulsen et al. 2003 for reviews).
These interactions are creating X-ray surface brightness depressions
or cavities that, like bubbles in soda water,  move buoyantly through
the intracluster medium.  The bubbles in some (but not all) systems,
contain enough energy to balance  radiative losses emerging from
the centers of clusters in the
X-ray band.  This, along with the discovery of very short
central cooling timescales, and at the same time,
the lack of evidence for strong cooling below X-ray temperatures, 
have renewed interest
in feedback-driven heating mechanisms capable of balancing radiation
losses (Nulsen, this conference).
It is now an established fact that the keV gas in clusters
with cooling times approaching 100 Myr are frequently associated with
the sites of star formation.  Therefore,
this star formation may have been fueled during 
the cooling phase of the feed-back loop.

\section{Properties of Star Formation in Clusters}

I list below a few of the key
observational facts about star formation in cooling flows.
(I use the term ``cooling flow'' to refer to 
clusters with short central cooling times.)
For more detailed discussions and additional references,
see recent reviews by B\"ohringer et al. (2001), 
Crawford (2003 \& this conference), Fabian (1994), 
and McNamara (1997, 2002).

\begin{enumerate}

\item A trend exists between cooling flows and
the occurrence and amplitude of blue color
excesses associated with star formation in central
cluster galaxies (Johnstone et al. 1987, McNamara \& O'Connell 1989, 
Cardiel et al. 1998, Crawford et al. 1999).  
{\it Chandra} has shown that the regions of star formation
are associated with bright lumps and filaments of gas whose 
radiative cooling times approach $\sim 10^8$ yr 
(McNamara et al. 2000, Fabian et al. 2001,
McNamara et al. 2004, Blanton et al. 2003).
Abell 1795, shown in Figure 1, is a good example (Fabian 2001).

\item Bright, spatially extended nebular emission
is seen preferentially in clusters (e.g. Perseus) with central cooling times
below  $\sim 1$ Gyr (Hu 1988; Heckman et al. 1989).   
Recent comparisons between optical emission line maps and
{\it Chandra} X-ray maps have
shown spatial correlations between nebular emission and
bright lumps and filaments of gas (Figure 1) where the radiative cooling
time approaches $\sim 10^{8}$ yr (Fabian et al. 2003, Blanton et al. 2001).

\begin{figure}
\plotone{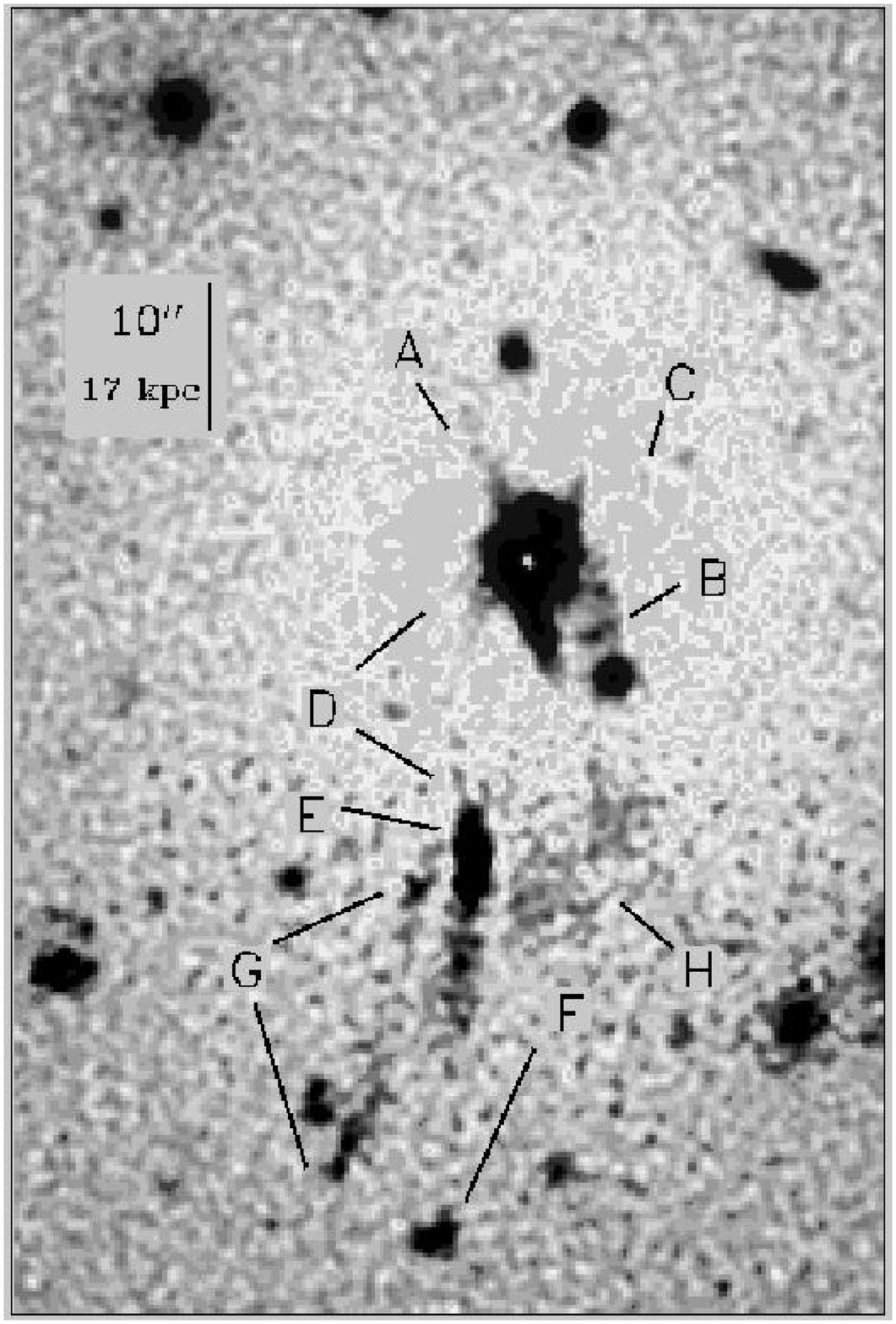}
\end{figure}

\begin{figure}
\plotone{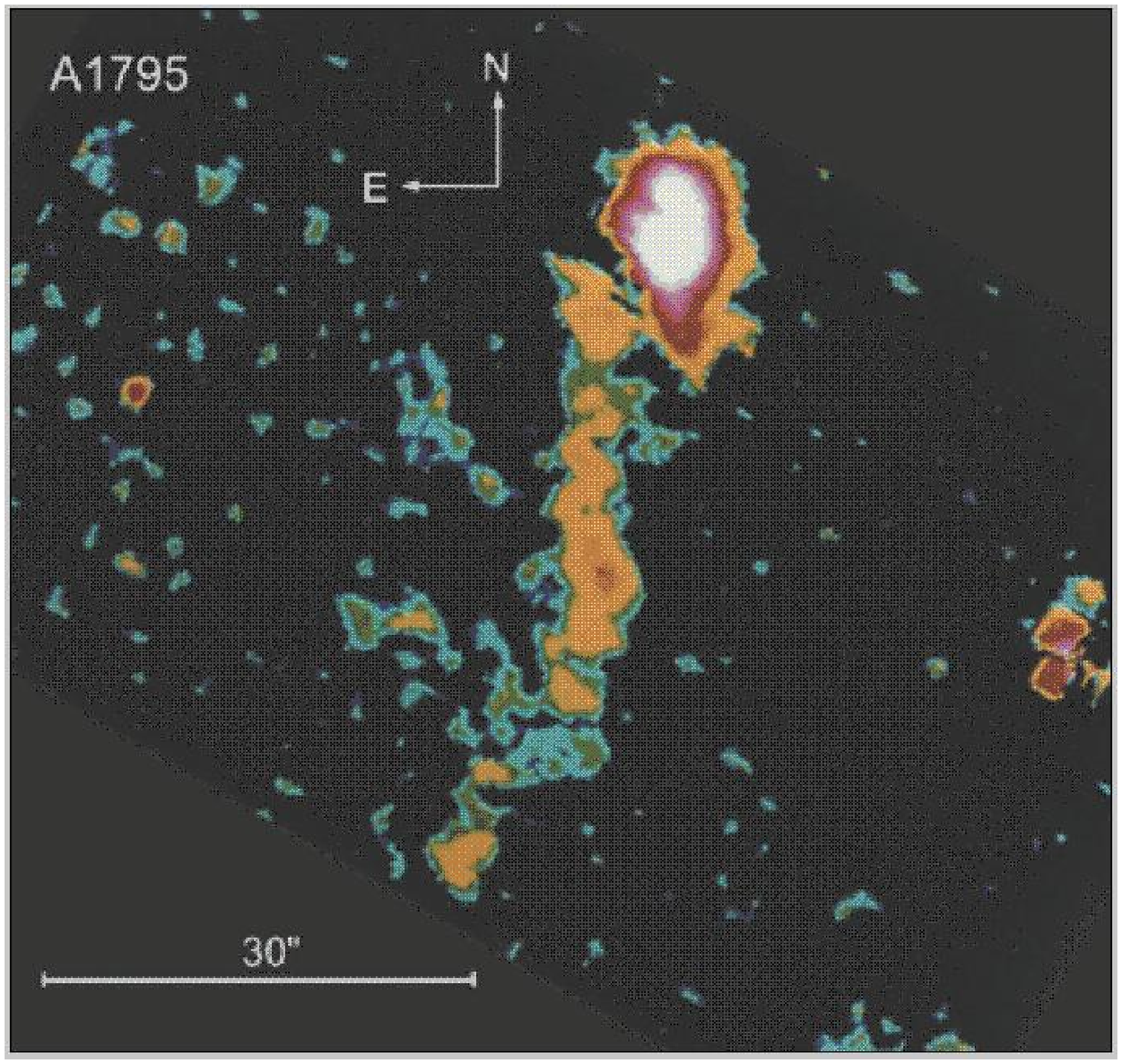}
\end{figure}

\begin{figure}
\plotone{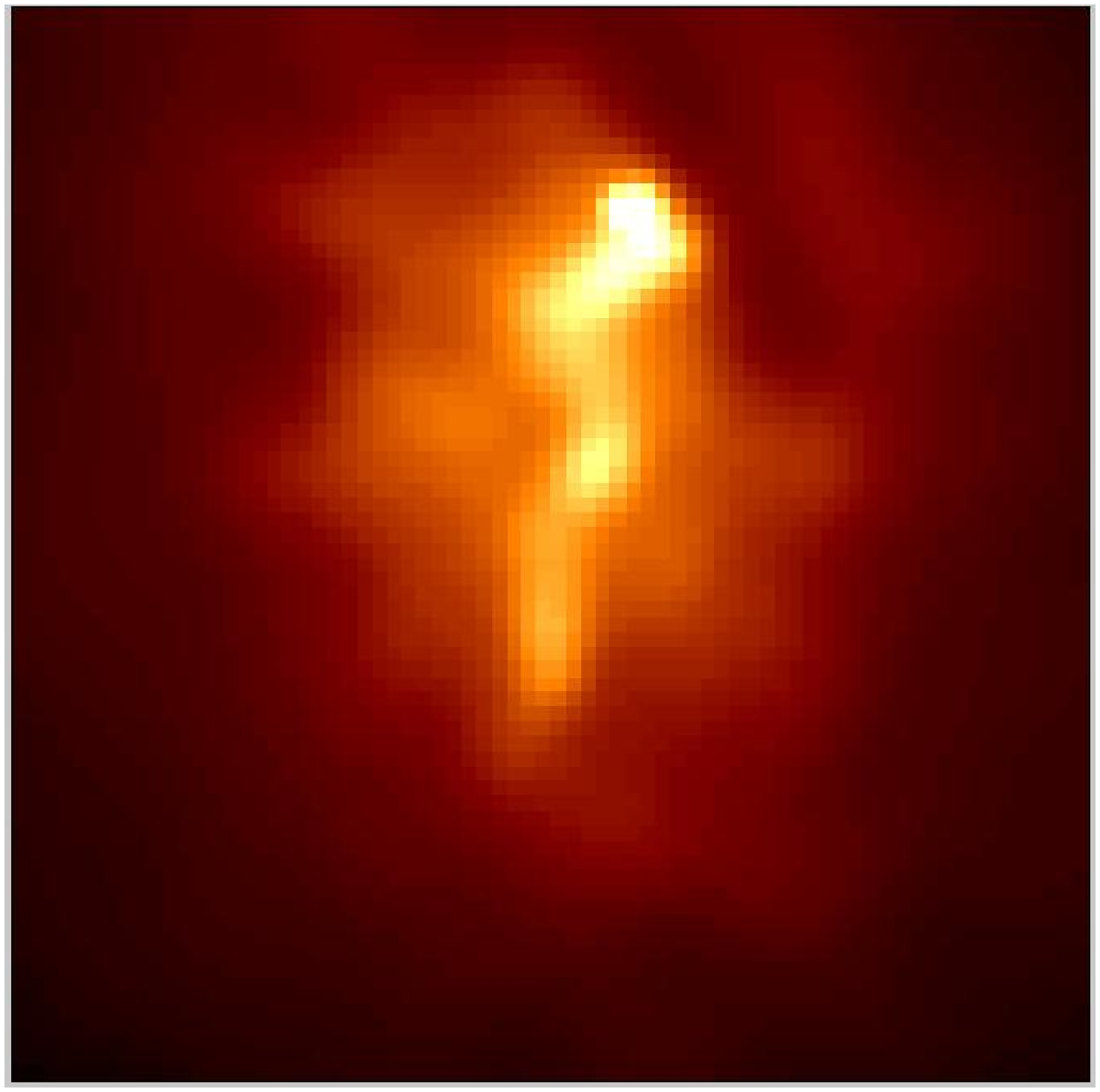}
\figcaption{{\bf Upper Left:} Star formation map (McNamara et al. 1997),
{\bf Upper Right:} H$\alpha$ map (Cowie et al. 1983), and {\bf Lower Right:}
Chandra X-ray map of the center of the Abell 1795 cluster (Fabian et al. 2001).  The trail of
star formation, indicated by lettering, is traced using a processed,
deep U-band image (McNamara et al. 1997).  The U-band, H$\alpha$,
and X-ray trails are clearly correlated.
\label{Scott:myplot}}
\end{figure}

\item The star formation histories vary between short duration bursts 
$\lae 10$ Myr of age, or  more extended episodes lasting
between 0.1-1 Gyr; such histories are inconsistent with
steady cooling and accretion that has endured for the 
ages of clusters (Allen 1995, Crawford et al 1999, McNamara 1997, McNamara
et al. 2004).  The short duration bursts are, in some cases, triggered by
the central radio source (McNamara \& O'Connell 1993).

\end{enumerate}

\section{A Comparison Between the Cooling and Star Formation Rates}

A clean demonstration of equality between the level of
cooling and star formation plus its associated gas would be a critical
test of the new, self-regulated cooling paradigm.
More importantly, if it can be shown 
that the cooling upper limits are systematically below
the star formation levels, a possibility that is now within reach,
we would be in a position to reject the link between cooling 
and star formation, with some measure of confidence. 

Cooling rates have now been estimated for several clusters
with {\it Chandra} and XMM-{\it Newton}.  
As I discussed above, they are systematically
below the classical values. 
This trend is shown in Figure 2, where I plot cooling rate ($\dot M$) versus 
star formation rate (SFR), in solar masses per year, for five clusters.
The solid squares show the morphological cooling rates
from the {\it ROSAT/Einstein} era,  calculated
essentially by dividing the central gas mass
by the cooling time.  The modern cooling rates fall well below these
values and are shown along with their measurement uncertainties.  Only
Abell 1795 and Abell 2597 have independent cooling measurements
from {\it FUSE} spectra of the O VI $\lambda 1032$
feature (Oegerle et al. 2001).  In both cases, the ultraviolet cooling
rates, shown as filled dots in Figure 2, 
lie within the uncertainties of the X-ray cooling rates, 
represented as vertical bars.   The cooling rates for
the remaining objects are presented as upper limits because
the {\it Chandra}-ACIS spectral cooling rates represent the ``maximum cooling''
models consistent with the data (Wise et al. 2004), and they
should not be misconstrued as the unique spectral signatures 
of cooling below one or
two keV. In other words, in spite of the very short cooling time of the 
keV gas, single temperature model plasmas provided acceptable fits
to the data at each radius.

The star formation rates owe their uncertainties (horizontal
bars)
primarily to the difficult problem of measuring the intrinsic colors
of young accretion populations against the background
glare of the cD galaxy, mismeasuring extinction, and degeneracies in the
population models themselves (cf., McNamara et al. 2004).  

\begin{figure}
\plotone{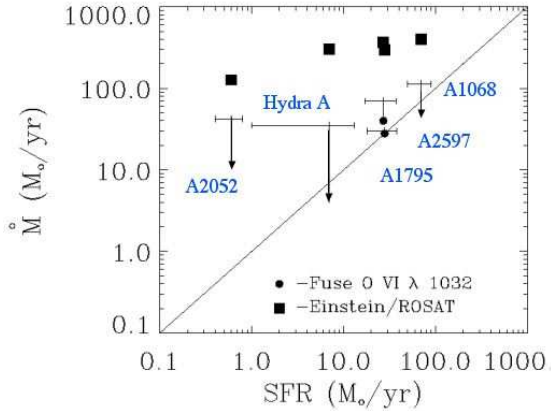}
\figcaption{Plot of X-ray cooling rate versus star formation rate. The 
line represents equality between the rates. An
explanation of this figure is given in the text.
\label{Scott:myplot}}
\end{figure}

Figure 2 shows that the cooling and star formation rates are generally
converging, and in some cases they agree to within the (substantial)
measurement errors.  This is evident in Abell 1068,
Abell 2597, and Abell 1795.  While I don't regard this as proof positive
that the cooling flow problem has been solved,
we are clearly on the right track.  The challenges ahead include
improving our understanding of the star formation rates and 
histories, and obtaining higher precision (deeper) X-ray spectra capable
of turning the current upper limits into cooling detections, or excluding
cooling entirely at levels below the the star formation rates.  

\section{What is Preventing Most of the Gas from Cooling?}

Essentially all indicators are telling us
that the enormous radiation losses in cluster cores 
are balanced, or nearly so, by some form of heating.  
Clear-cut evidence now exists that interactions between 
radio sources and the surrounding gas are supplying 
some of this heat in many clusters and perhaps all
of it in others (see Nulsen and Blanton this conference).  
Cavities have been identified in nearly two dozen clusters 
over the past three
years (B\^{\i}rzan et al. 2004).   The archetypes, Hydra A (McNamara
et al. 2000) and Perseus (B\"ohringer et al. 1993, Fabian et al. 2000),
are typical of most systems: 
twin surface brightness depressions $10-20$ 
kpc in diameter lying at distances of $10-30$ kpc from the nucleus of the cD.
Cavities have also been observed in giant elliptical galaxies,
such as M84 (Finoguenov \& Jones 2001), and groups, such as HGG 62
(B\^{\i}rzan et al. 2004).  Cavity ages range between  
$\sim 10^7~{\rm yr}-10^8~{\rm yr}$. Their enthalpy
ranges between $\gamma pV/(\gamma -1)\sim 10^{55}~{\rm erg}$ in
isolated galaxies and groups to $\sim 10^{60}~{\rm erg}$
in rich clusters.  The total energy input from each AGN outburst
may be several times these figures  when shocks are included (Fabian 2003,
Forman et al. 2004).  
The total energy deposited into the intracluster medium integrated 
over the lifetime of the AGN can greatly exceed $10^{61}$ erg (McNamara
et al. 2001, and see Nulsen et al. 2003 for a review). 
Nevertheless, while AGN may be able to retard or quench cooling in 
many systems, they would do so with great difficulty in others.
Only  $\sim 25\%$ of clusters in the {\it Chandra} archive
have obvious cavity systems, and many of these systems contain
too little energy to balance radiative losses at the current time
(B\^{\i}rzan et al 2004).

Other heating agents could be assisting the AGN at balancing
radiative losses.  Of these, heat conduction between the cool cores and hot outer layers of clusters has received a good deal 
of attention (Voigt et al. 2002).  But this model
has problems. In order to work effectively,
most studies have concluded that conduction must be suppressed by
several times the Spitzer rate.  While this may be
true in special cases, it cannot be easily demonstrated
observationally.  In other systems, 
heat conduction acting alone at the Spitzer rate
is incapable of balancing radiative losses 
(Voigt et al. 2002, Zakamska \& Narayan 2003, Wise et al. 2004).
Additional agents acting together to a greater or lesser degree,
such as mergers (Motl et al. 2003) and
supernovae associated with star formation (McNamara et al. 2004), 
must be assisting the AGN.

\section{Conclusions \& Speculations about a New Cooling Flow Paradigm} 

I have shown that the star formation rates estimated with optical,
ultraviolet, and infrared observations are within factors of
several of the new cooling limits, and in some cases, 
agree to within their errors.  What we know about 
the star formation histories in cooling flows points to repeated
bursts of star formation lasting $10^7 -10^9$ yr.  
In some cases, the starbursts are triggered by the central radio sources.
Constant
star formation over the ages of clusters, the history of star formation
predicted by steady cooling flow models, is inconsistent with most
data. 

The overall picture of star formation in cooling flows 
is consistent with, and indeed has helped to shape, the emerging paradigm 
of self-regulated cooling in clusters.
The X-ray, optical, and radio data taken together point to  
episodes of cooling and periodic reheating by radio outbursts and
their associated bubbles.  This scenario (e.g., Churazov et al. 2002) 
has received observational
support from the newly-discovered trend between central X-ray
luminosity and the instantaneous kinetic luminosity of
bubbles in clusters (B\^{\i}rzan et al. 2004).  
Thermal conduction may be
an additional element of the feed-back loop that regulates cooling
(Ruszkowski \& Begelman 2002, Nulsen 2003). 
This emerging picture of cooling flows has broad implications
for theories of structure formation and evolution.  Feed-back processes
may have been important during the early development of galactic
bulges and their central black holes, and they 
may regulate the thermal balance
of the hot gas in giant elliptical galaxies today (Nulsen, this
conference).  The possibility that such processes are active
in large, bright, relatively nearby clusters provides
a unique opportunity to test these models in detail.



\acknowledgements
I would like to acknowledge my colleagues Michael Wise, Paul Nulsen, 
Craig Sarazin, Larry David, and Chris Carilli, and my graduate
students  Laura B\^{\i}rzan and David Rafferty for their contributions to the 
work discussed here.  I would also like to thank the Local Organizing
Committee, Thomas Reiprich in particular, for hosting a great meeting,
and Noam Soker for making it happen. 
This research is supported by generous grants
from NASA, the Chandra X-ray Center, the Space Telescope Science
Institute, and the Department of Energy, including LTSA grant NAG5-11025
and Chandra General Observer and Archival Research Awards GO0-1078A, 
AR2300-7X, and GO1-2139X.


\end{document}